\documentclass[preprint,aps,superscriptaddress,floatfix]{revtex4}
\usepackage{graphicx}
\usepackage{xcolor}
\usepackage{amsmath}
\usepackage{hyperref}
\begin{document}
\title[Plasma Oscillator]{Energy Transmission and Harmonic Propagation in Unidirectionally Coupled Plasma Oscillator Networks}

\author{K. Abirami}
\email{nk.abirami@gmail.com}
\affiliation{Centre for Nonlinear and Complex Networks, SRM Institute of Science and Technology, Ramapuram, Chennai 600089, Tamil Nadu, India}
\author{Miguel A. Prado-Reynoso}
\email{miguelangel.prado@urjc.es}
\affiliation{Nonlinear Dynamics, Chaos and Complex Systems Group, Departamento de F\'{i}sica, Universidad Rey Juan Carlos, Tulip\'{a}n s/n, 28933 M\'{o}stoles, Madrid, Spain}
\author{Miguel A.F. Sanjuán}
\email{miguel.sanjuan@urjc.es}
\affiliation{Nonlinear Dynamics, Chaos and Complex Systems Group, Departamento de F\'{i}sica, Universidad Rey Juan Carlos, Tulip\'{a}n s/n, 28933 M\'{o}stoles, Madrid, Spain}
\affiliation{Royal Academy of Sciences of Spain, Valverde 22, 28004 Madrid, Spain}

\date{\today}

\begin{abstract}
We investigate signal transmission in a network of unidirectionally coupled plasma oscillators, where only the first oscillator is driven by a weak periodic input. As the signal propagates through the network, nonlinear interactions generate higher-order harmonic components. Spectral analysis identifies five distinct transmission patterns for the fundamental frequency and odd harmonics, and four patterns for even harmonics, encompassing damped propagation, undamped transmission, and localized amplification. The transmission characteristics are governed by both the coupling strength and system parameters. Furthermore, we analyze how variations in these key parameters shape the overall signal dynamics.
\end{abstract}

\maketitle
\section{Introduction}

Plasma oscillations have attracted significant attention in recent years due to their fundamental role across a wide spectrum of scientific and technological domains. As intrinsic manifestations of collective plasma behavior, these oscillations underpin numerous plasma-based technologies, including semiconductor fabrication processes \cite{chen2015introduction}, surface-modification techniques \cite{lieberman1994principles}, biomedical instrumentation \cite{kong2009plasma}, and environmental control systems \cite{tendero2006atmospheric}. Beyond their practical relevance, a central objective in plasma physics is to understand the nonlinear behaviors and instabilities that govern energy transport, wave–particle interactions, and dynamic plasma control. Investigations into such nonlinear plasma dynamics have proven essential for advancing materials science, high-precision manufacturing, and emerging energy technologies \cite{chian2022nonlinear,shiratani2011nano}.

Forced plasma oscillations, in particular, play a pivotal role in applications ranging from plasma heating and diagnostics to the study of nonlinear wave phenomena in both laboratory and industrial settings \cite{moulick2022fundamentals}. In magnetic confinement fusion devices such as tokamaks, external periodic forcing is indispensable for achieving the extreme conditions required for sustained nuclear fusion. Through careful tuning of the drive frequency and amplitude, it is possible to probe key plasma parameters, such as density, temperature, and magnetic field strength, while also exerting control over energy transport and signal propagation \cite{hebner2000overview, samikkannu2023studies, liu2019high}. Nonlinear plasma processes such as electron trapping, detrapping, and double-layer formation have been experimentally observed to induce complex wave-wave interactions and instabilities in DC glow discharge plasmas \cite{singh2024trapping}.

To model these driven systems, researchers have developed simplified yet effective dynamical frameworks that capture the essential physics of nonlinear plasma responses. A notable example is the nonlinear anharmonic oscillator model introduced by Enjieu-Kadji and collaborators, which describes plasma oscillators subjected to periodic external forcing \cite{kadji2008regular}. This approach successfully replicates fundamental phenomena such as amplitude modulation, frequency locking, and nonlinear resonance. Under certain parametric conditions, the system exhibits rich dynamical behavior, including bifurcations, mixed-mode oscillations, and transitions to chaos \cite{chen2024bifurcations}. A similar model has been employed to describe the fluctuating potential function observed in a DC glow discharge plasma system with a constricted anode geometry \cite{saha2014investigation}. The experimental work by Jayaprakash et al. \cite{jayaprakash2021doubly} provides empirical evidence supporting the anharmonic plasma oscillator framework, offering important insights into nonlinear behaviors observed in plasma systems. Expanding upon this framework, further studies have revealed that modulated external fields can trigger subtle resonance effects such as ghost-vibrational resonance and enhanced vibrational amplification especially in multi-fluid plasma models \cite{samikkannu2023studies, bhuvaneshwari2022enhanced}. These findings underscore the system’s heightened sensitivity to external perturbations in the nonlinear regime. Recent theoretical developments in aperiodic resonance phenomena which involve irregular yet coherent responses to nonperiodic or complex external driving have also provided new insights into how such mechanisms operate in nonlinear systems, including plasma environments \cite{yang2025aperiodic}. 

Understanding energy transport in unidirectionally coupled nonlinear networks is of growing importance in both theoretical and applied physics\cite{zhu2020unidirectional,yamamoto2016unidirectional}. In these networks, each oscillator is influenced only by its immediate predecessor, establishing a directional, feedforward flow of energy \cite{yao2010signal}. This setup closely resembles real-world systems where energy or information is injected at one end and propagates through a nonlinear medium. Analytical studies have reported stationary one-dimensional solutions describing the coupled propagation of nonlinear electron plasma and ion acoustic waves, where the amplitudes are linearly proportional and propagate as periodic wave trains with ion-acoustic velocity, including solitary waves as a limiting case \cite{nishikawa1974coupled}. Synchronization between two coupled plasma were studied experimentally in \cite{chaubey2015synchronization}.

Unidirectional coupling has been studied in various contexts of nonlinear dynamics, where the coupling strength is an important control parameter that influences the collective behavior of oscillator networks \cite{stankovski2017coupling}. In Ref.~\cite{yao2010signal}, Yao et al. investigated signal transmission in a unidirectionally coupled Duffing oscillator network under biharmonic forcing using the vibrational resonance approach. They identified two transmission regimes damped and undamped in both cases considered. Similarly, Rajamani et al.~\cite{rajamani2013signal} studied signal amplification in unidirectionally coupled Duffing oscillators driven by a monoharmonic force and reported damped and undamped signal transmission. In addition, they demonstrated stochastic amplification in network nodes by explicitly introducing randomness through randomly connected or randomly excited oscillators in a two-dimensional network of Duffing oscillator.

Another related study by Tchakui et al.~\cite{tchakui2023dynamics} examined unidirectionally coupled non-identical Van der Pol oscillators driven by a sinusoidal signal at the first node. For identical oscillators, the transmitted signal amplitude remained constant along the network, whereas non-identical oscillators exhibited amplitude fluctuations that were progressively suppressed with increasing coupling strength. These fluctuations arise from system non-identicality and are further enhanced by the presence of cubic nonlinearity. It is worth noting that, although the Van der Pol oscillator involves nonlinear damping, its dynamical structure differs fundamentally from the plasma oscillator considered in the present work.

When plasma oscillators are arranged in a unidirectionally coupled topology, the resulting dynamics can give rise to novel emergent phenomena such as nonreciprocal energy transmission, complex spatiotemporal signal evolution, and dynamical pattern formation. These behaviors raise several fundamental questions: How far can a coherent signal travel through a nonlinear network? What factors control the amplification, attenuation, or localization of signal? And how do nonlinear interactions and external forcing influence the structure of signal propagation?

This study addresses a significant gap in understanding how unidirectional coupling influences resonance behavior, higher harmonic generation, and the overall efficiency of signal transmission in nonlinear oscillator networks. The insights gained contribute to the broader goal of controlling wave-based signal transport in plasma and support the development of advanced technologies in areas such as plasma diagnostics\cite{thomas2024plasma}, RF communication\cite{pena2025flexible}, and neuromorphic computing \cite{kipelkin2023mathematical}.

In this work, we investigate the nonlinear resonance behavior and signal transmission dynamics of single, ten, and 100 plasma elements, each modeled as a driven nonlinear plasma oscillator. The focus is on unidirectional coupling to understand how signals propagate and evolve under the combined effects of nonlinearity, damping, and interaction strength. Section 2 provides a detailed description of the governing model. Section 3 presents both numerical and analytical analyses of unidirectionally coupled plasma oscillators. Section 4 discusses signal transmission in a network of $n$ unidirectionally coupled plasma oscillators. Section 5 extends the framework by examining amplitude spectra, higher harmonic generation, and spatiotemporal signal evolution across the network. A systematic parametric study is also carried out in Section 5 to evaluate the influence of damping, asymmetry, and cubic nonlinearity on energy transport. Finally, Section 6 summarizes the key findings and highlights their broader implications.

\section{Model: unidirectional network of $n$ Plasma oscillator}

In this work, we focus on the two-fluid plasma oscillator model, which effectively captures the nonlinear dynamics of plasma type systems by treating ions and electrons as interpenetrating fluids governed by Maxwell’s equations. Derived from the Euler equations with electromagnetic coupling and source terms, this model provides a robust framework for analyzing plasma instabilities, such as fluctuations in charge density, electric potential, and temperature, as well as complex phenomena including nonlinear resonance, bifurcations, and signal transmission.

Following \cite{enjieu2008nonlinear}, the governing equation for an isolated, non-dimensional, driven plasma oscillator with potential $V(x)$ is given by
\begin{equation}
\ddot{x} + d(1 + x^2)\dot{x} + V'(x) = f \cos(\omega t), \label{main}
\end{equation}
where the over dot and prime denote derivatives with respect to time and space, respectively. The variable $x(t)$ represents the displacement of a localized plasma element \cite{merlino2021dusty}. The parameter $d$ is the nonlinear damping coefficient, $V(x)$ denotes the confining potential, and $f\cos(\omega t)$ is an external periodic forcing with amplitude $f$ and angular frequency $\omega$. Physically, $x$ may be interpreted as the displacement of a charged dust grain in a dusty plasma or of an electron–ion element in a bounded plasma system.

\begin{figure}[htbp]
\begin{center}
\includegraphics[height=0.3\textwidth,width=0.35\textwidth]
{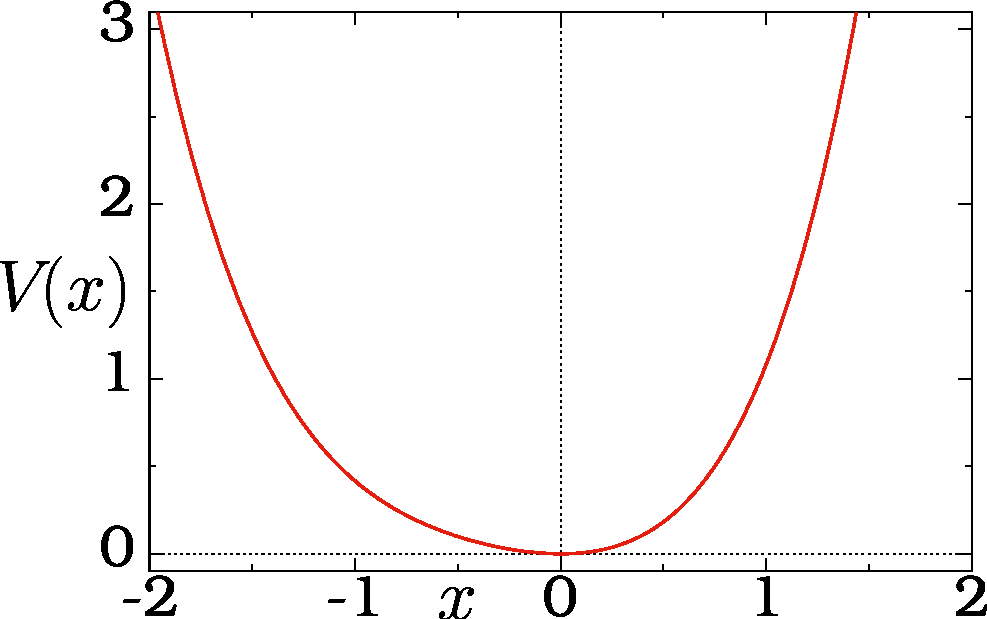}
\end{center}
\caption{Asymmetric potential well of the plasma oscillator with parameters $\omega_0^2=1,~\alpha=1,$ and $\beta=1.$}
\label{figure1}
\end{figure}
To capture the asymmetry essential for modeling realistic plasma type responses under directional excitation, the system evolves under an asymmetric anharmonic potential\cite{jeyakumari2011vibrational},
\begin{equation}
V(x) = \frac{\omega_0^2}{2} x^2 + \frac{\alpha}{3} x^3 + \frac{\beta}{4} x^4,
\end{equation}
where $ \omega_0^2 $ is the linear frequency parameter, and $ \alpha $ and $ \beta $ correspond to the coefficients of quadratic and cubic nonlinearities, respectively. In this work, we consider an asymmetric potential featuring a single minimum, as illustrated in Fig. \ref{figure1}. The parameters are fixed at $\omega_0^2=\alpha=\beta=1$.

\begin{figure}[htbp]
\begin{center}
\includegraphics[height=0.12\textwidth,width=0.7\textwidth]
{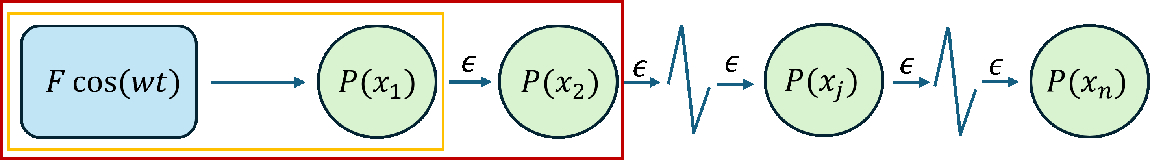}
\end{center}
\caption{Schematic representation of a unidirectionally coupled network of $n$ identical plasma oscillators, with coupling strength $\epsilon$. The system is externally driven by a periodic force applied only to the first oscillator. The yellow block represents an isolated driven plasma oscillator, while the red block highlights a pair of unidirectionally coupled plasma oscillators.}
\label{figure2}
\end{figure}

To model directional interactions observed in many plasma environments, we extend the system to a network of $n$ identical plasma oscillators with unidirectional coupling. Each oscillator interacts only with its immediate predecessor, and the network is externally driven by a periodic force applied solely to the first oscillator. This configuration mimics directional energy or signal propagation found in real plasma systems. A schematic representation is shown in Fig.~\ref{figure2}. The yellow block represents an isolated driven plasma oscillator, while the red block illustrates two unidirectionally coupled plasma oscillators.

The governing equations describing the dynamics of this configuration are given by:
\begin{eqnarray}
\ddot{x}_1 + d(1+x_1^2) \dot{x}_1 + \omega_0^2 x_1 + \alpha x_1^2+\beta x_1^3 &=& f \cos(\omega t)\nonumber \\
\ddot{x}_i + d(1+x_i^2) \dot{x}_i + \omega_0^2 x_i + \alpha x_i^2+\beta x_i^3 \ &=& \epsilon x_{i-1},\quad \qquad i=2,3,\dots n. \label{network_eqn}
\end{eqnarray}

where $\epsilon$ characterizes the unidirectional coupling strength. The analysis is carried out by computing both numerical solutions of the governing equation and analytical approximations obtained via the method of harmonic balance. This combined approach enables the identification of key features, such as resonance peaks, nonlinear frequency shifts, and the influence of potential asymmetry on the overall response.
\section{Analysis of the Frequency Response of Plasma Oscillators Under Unidirectional Coupling Strength}
To investigate the system's resonance dynamics, all oscillators are initially set at the equilibrium state $(0,0)$. The first oscillator is then driven into a stable steady state oscillation around the equilibrium, well away from any complex dynamical regime, while all oscillators are chosen to be identical, with parameters $\alpha= \beta= \omega_0^2=1.0$ and $d=0.5$. Through the unidirectional coupling, the oscillatory response of the first oscillator is transmitted to the subsequent elements, thereby inducing oscillations along the network. For the numerical simulations, Eq.~(\ref{network_eqn}) is integrated using the classical fourth order Runge Kutta method with a fixed time step of $h=0.01$. 

After computing the time-domain solution, the steady-state response of the oscillator at the excitation frequency $ \omega $ is extracted using the Fourier sine and cosine transform. For each element of Eq.~(\ref{network_eqn}), the response amplitude $ Q_i(\omega) $ is computed as
\begin{equation}
Q_i\left(\omega\right) = f^{-1} \sqrt{Q_{i}^s\left(\omega\right)^2 + Q_{i}^c\left(\omega\right)^2},
\end{equation}
where:
\begin{subequations}
		\label{e8}
		\begin{equation}
			Q_i^s (\omega)= \frac{2}{NT} \int_{T_0}^{T_0 + NT} x_i \sin(\omega t) \, dt,
		\end{equation}
		\begin{equation}
			Q_i^s (\omega)= \frac{2}{NT} \int_{T_0}^{T_0 + NT} x_i \cos(\omega t) \, dt,
		\end{equation}
\end{subequations}
with \(T = 2\pi / \omega\), correspond to the normalized Fourier sine and cosine coefficients of the steady-state response, computed over $N=200$ periods to extract the amplitude at the driven frequency $\omega$. To eliminate the influence of transient dynamics, a sufficiently long transient interval of $T_0=1000$ is discarded before analysis.

As a first approximation, each plasma oscillator is assumed to undergo a steady–state harmonic oscillation at the driving frequency $\omega$, while higher harmonics and transient contributions are neglected. Accordingly, the response of the $i$-th oscillator is written as
\begin{equation}
x_i(t)=A_i\cos(\omega t+\phi_i), \qquad i=1,\dots,n,
\end{equation}
where $A_i$ and $\phi_i$ denote the response amplitude and phase shift, respectively. Substituting this trial solution into Eq.~(\ref{network_eqn}) and applying the harmonic balance method \cite{rajasekar2016nonlinear,jothimurugan2016multiple}, we obtain
\begin{eqnarray}
&&\dot{x}_i=-A_i\omega\sin(\omega t+\phi_i),\nonumber \\
&&\ddot{x}_i=-A_i\omega^2\cos(\omega t+\phi_i),\nonumber\\
&&x_i^3=A_i^3\cos^3(\omega t+\phi_i)
=\frac{A_i^3}{4}\Big[\cos(\omega t+\phi_i)+\cos(3\omega t+3\phi_i)\Big],\nonumber\\
&&x_i^2=\frac{A_i^2}{2}\Big[1+\cos(2\omega t+2\phi_i)\Big],\nonumber\\    
&&x_i^2\dot{x}_i=-A_i^3\omega\sin(\omega t+\phi_i)\cos^2(\omega t+\phi_i)
=-\frac{A_i^3\omega}{4}\Big[\sin(\omega t+\phi_i)+\sin(3\omega t+3\phi_i)\Big], \nonumber\\
&&f\cos(\omega t)=f\cos(\omega t+\phi_i-\phi_i)
=f\Big[\cos\phi_i\cos(\omega t+\phi_i)+\sin\phi_i\sin(\omega t+\phi_i)\Big].
\end{eqnarray}
Substituting these expressions into Eq.~(\ref{network_eqn}), retaining only the fundamental harmonic terms, and equating the coefficients of $\cos(\omega t+\phi_i)$ and $\sin(\omega t+\phi_i)$, we obtain
\begin{eqnarray}
A_i\!\left(\omega_0^2-\omega^2+\frac{3\beta A_i^2}{4}\right) &=& f\cos\phi_i,
\label{eq7}\\
-A_i\!\left(1+\frac{A_i^2}{4}\right)d\omega &=& f\sin\phi_i .
\label{eq8}
\end{eqnarray}
Squaring and adding equation \ref{eq7} and \ref{eq8} yields
\begin{eqnarray}
&&A_i^2\!\left(\omega_0^2-\omega^2+\frac{3\beta A_i^2}{4}\right)^2
+A_i^2\!\left(1+\frac{A_i^2}{4}\right)^2 d^2\omega^2 -F_i^2 = 0, \\
&&\phi_i = \tan^{-1}\!\left(\frac{-\left(1+\frac{A_i^2}{4}\right)d\omega}
{\omega_0^2-\omega^2+\frac{3\beta A_i^2}{4}}\right),
\label{eq5}
\end{eqnarray}
where $F_1^2=f^2$ and $F_i^2=\epsilon^2 A_{i-1}^2$ for $i=2,3,\ldots,n$.

Equation~(\ref{eq5}) provides a nonlinear relationship between the response amplitude and the driving frequency. It is solved numerically using the Newton–Raphson method, and the normalized analytical response amplitude is then evaluated as $Q_i(\omega)=A_i/f$.

Due to the system’s unidirectional interaction, we systematically investigate its frequency response, beginning with a single forced element and progressively extending the analysis from a few coupled elements to a network of $100$.

\subsection{Isolated Driven Forced Plasma oscillator}
We first examine the steady-state frequency response of a single externally driven plasma oscillator, as a baseline for understanding the more complex dynamics that emerge in coupled oscillator networks.
\begin{figure}[h]
\begin{center}
\includegraphics[height=0.4\textwidth, width=0.5\textwidth]
{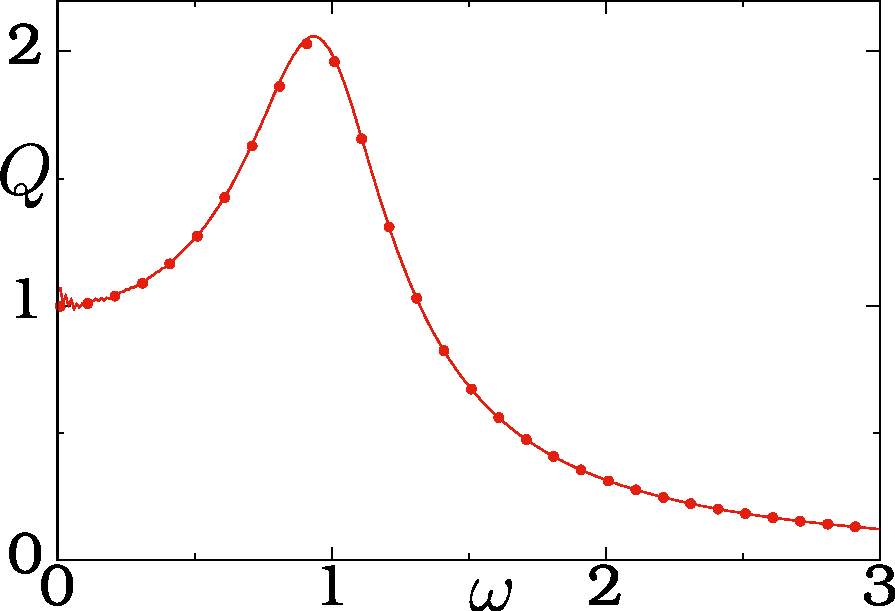}
\end{center}
\caption{Frequency response of the isolated plasma oscillator. The solid line represents the numerically computed amplitude $Q$, while the dots indicate the analytically calculated values of $Q$.}
\label{figure3}
\end{figure}

Figure~\ref{figure3} shows the response amplitude of the single driven plasma oscillator with driving frequency $\omega$. The numerically computed values of $Q$ are plotted as a continuous line, while the analytically predicted values are represented by dots. A good agreement is observed between the numerical and analytical results. As $ \omega $ increases from a low value $ \omega = 0.01 $, the response amplitude initially increases, indicating that the plasma oscillator is progressively absorbing energy from the external forcing and undergoes oscillations. The response amplitude reaches a maximum value of $ Q = 2.06 $ at the resonance frequency $ \omega = 0.94 $, demonstrating a classical resonance phenomenon in which the plasma oscillator exhibits maximum energy transfer from the external drive. Beyond this point, the amplitude decreases gradually with further increases in $ \omega $, reflecting the diminishing efficiency of energy absorption at higher frequencies. This characteristic frequency response highlights the nonlinear resonance profile of the plasma oscillator under single-frequency excitation.

\subsection{Network of 10 plasma oscillators}
In this section, we analyze the collective response of a network of ten unidirectionally coupled oscillators, focusing on how the input signal propagates and amplifies across the array. The governing Eq.~(\ref{network_eqn}) is solved for $n=10$, with parameters fixed as $\alpha=\beta=\omega_0^2=\epsilon=1.0, \,d=0.5$ and $f=0.05$. These values place the system in a weakly nonlinear regime, enabling clear observation of resonance behavior and signal-transmission characteristics.

Under these conditions, the coupling direction determines the flow of information: each oscillator receives the output of its immediate predecessor as an input, creating a forward transmission path from the first oscillator ($i=1$) to the last ($i=10$). Such unidirectional coupling typically leads to cumulative amplification, where signal transferred from the driving oscillator is progressively reinforced through successive units in the network.
\begin{figure}[h!]
\centering
\begin{minipage}{0.35\textwidth}
\centering \includegraphics[width=\textwidth,height=\textwidth]{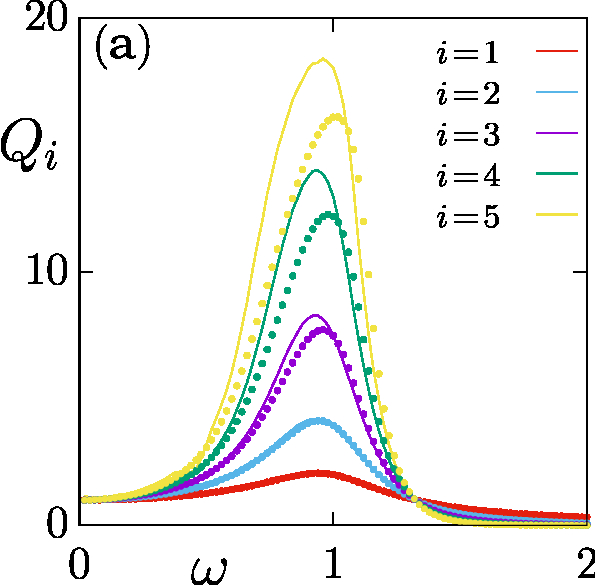}
\end{minipage}
\centering
\begin{minipage}{0.35\textwidth}
\centering \includegraphics[width=\textwidth,height=0.95\textwidth]{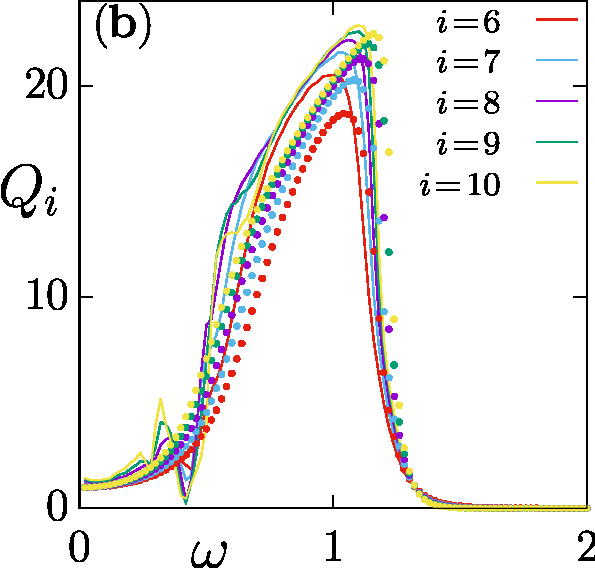}
\end{minipage}
\centering
\begin{minipage}{0.35\textwidth}
\centering \includegraphics[width=\textwidth,height=\textwidth]{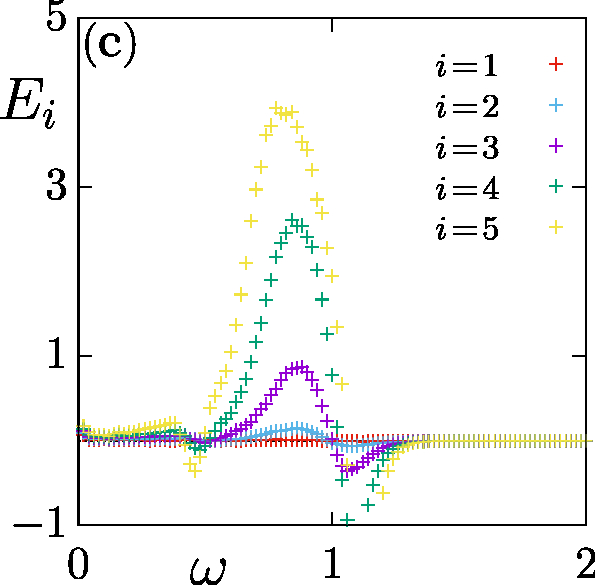}
\end{minipage}
\centering
\begin{minipage}{0.35\textwidth}
\centering \includegraphics[width=\textwidth,height=\textwidth]{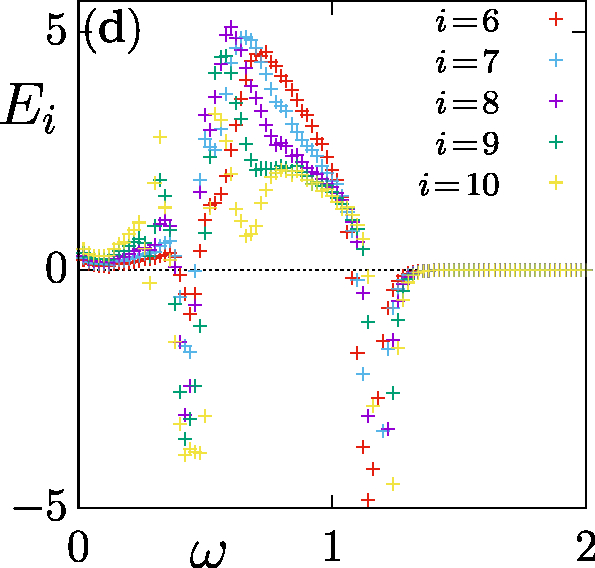}
\end{minipage}
\caption{Comparison between analytical and numerical response amplitudes for a unidirectionally coupled network of ten oscillators. Panels (a) and (c) show the response amplitude $Q_i$ as a function of the driving frequency $\omega$ for oscillators $i=1–5$ and $i=6–10$, respectively. The solid lines represent numerically computed values, while the dotted lines correspond to analytical predictions. Panels (b) and (d) display the corresponding deviation $E_i$.}
\label{figure4}
\end{figure}

Figure~\ref{figure4} presents the nonlinear response characteristics of each oscillator. Panels (a) and (c) show the variation of the response amplitude $Q_i$ with the driving frequency $\omega$. The continuous curves denote the numerically obtained values of $Q_i$, while the dotted curves represent the analytical predictions derived from the approximate theoretical formulation. The corresponding error functions, defined as $E_i=Q_i^{numerical}-Q_i^{analytical}$, are plotted in Figs.~\ref{figure4}(b) and \ref{figure4}(d) to quantify the deviation between the two approaches.

The results reveal a pronounced resonance around $\omega=1$, which corresponds to the natural frequency of the individual oscillators. Within the frequency range $\omega\in[0.44,1.3]$, strong amplification is observed, signifying efficient transmission of the external signal along the coupled network. The amplitude of the response increases from the first oscillator to the last, confirming that unidirectional coupling promotes enhanced signal propagation. This amplification saturates for higher oscillator indices, suggesting that the coupling strength $\epsilon=1$ is sufficient to achieve near-complete transmission within ten nodes. 

In the frequency range $\omega\in[0.44:1.3]$, the response amplitudes of the first two oscillators obtained from both approaches are nearly identical, with only a minimal deviation in $E_i$. For the third oscillator, 
$E_i$ remains less than 1, while for oscillator indices greater than 3, 
$E_i$ exceeds 1. The origin of the discrepancy between theoretical and numerical results is discussed in Section V. For excitation frequencies 
$\omega>1.3$, both the analytical and numerical results show a rapid decay of 
$Q_i$, indicating suppression of resonance at higher frequencies. In this regime, the two approaches remain in close agreement, demonstrating the robustness of the theoretical approximation.

Overall, this analysis demonstrates that the ten-oscillator unidirectional array effectively transmits and amplifies weak input signals over a finite frequency window. The analytical and numerical results show strong correspondence at low and high frequencies, confirming the validity of the model in these regimes. Furthermore, we extend our study to a network of $100$ unidirectionally interacting plasma oscillators, and a comprehensive analysis of its dynamical behavior is presented in the following section.

\subsection{Network of $100$ plasma oscillators.}
\begin{figure}[h]
\centering
\begin{minipage}{0.35\textwidth}
\centering 
\includegraphics[height=\textwidth,width=\textwidth]{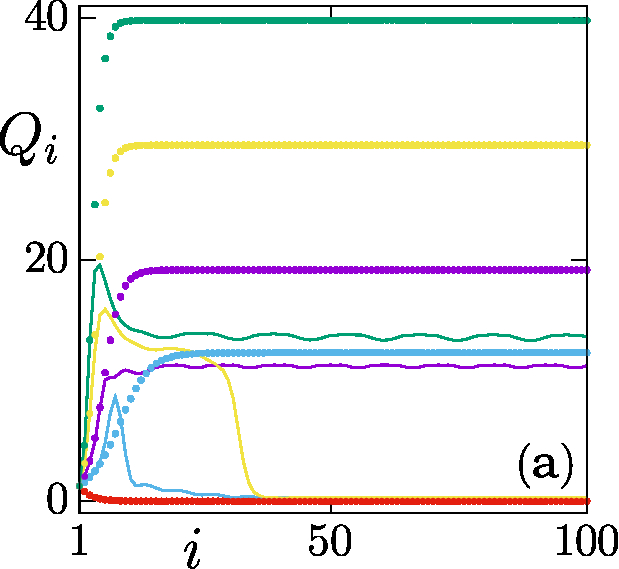}
\end{minipage}
\centering
\begin{minipage}{0.35\textwidth}
\centering 
\includegraphics[height=\textwidth,width=\textwidth]{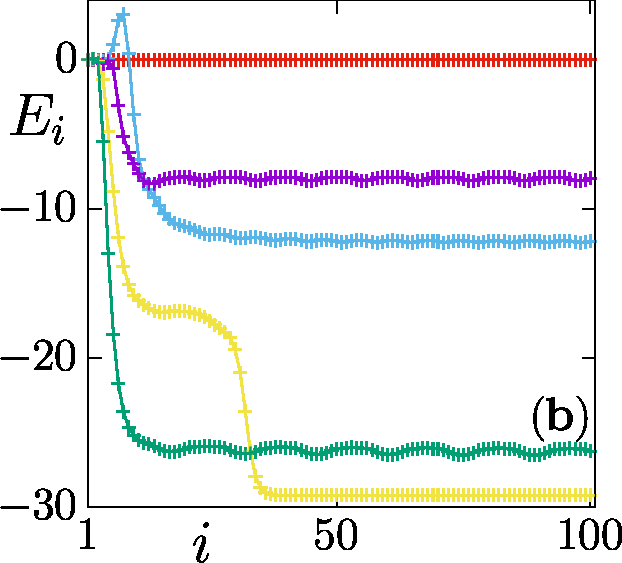}
\end{minipage}
\caption{(a) Variation of the response amplitude $Q_i$ with plasma oscillator $i$ for five different values of the coupling strength $\epsilon$, distinguished by color. The plus signs represent analytically calculated values of $Q_i$, while the continuous lines correspond to numerically computed results. (b) Difference plot of numerical and analytical $Q_i$'s in (a). Color coding is used to distinguish different values of the coupling strength $\epsilon$: red for $\epsilon = 0.5$, blue for $\epsilon = 1$, violet for $\epsilon = 1.3$, yellow for $\epsilon = 2$, and green for $\epsilon = 3$.}
\label{figure5}
\end{figure}

We now investigate signal transmission in a network of 100 unidirectionally coupled plasma oscillators. The system parameters and initial conditions are the same as in the previous section, except that the network size is increased from $n=10$ to $n=100$. In Fig.~\ref{figure5}, the analytical results are shown by symbols, while the continuous curves represent the numerically computed values of $Q_i$. For $\epsilon<0.75$, excellent agreement is observed between the analytical and numerical results, indicating a regime of damped signal transmission throughout the network. However, for  $\epsilon>0.75$, only the first few oscillators (typically the first four) continue to show good agreement, whereas noticeable discrepancies appear for the remaining elements. This deviation $E_i$ is clearly illustrated in Fig.~\ref{figure5}(b). While the analytical model predicts a monotonic undamped saturation of the transmitted amplitude for $\epsilon>0.75$, the numerical results reveal more complex transmission behaviors. The origin of this analytical-numerical mismatch is discussed in Section V. 
In the subsequent sections, we focus on analyzing the dynamics of the $ n $-plasma oscillators using numerical methods.
\section{Signal Transmission in a Network of $n$ Unidirectionally Coupled Plasma Oscillators}
To investigate signal transfer among $n$-plasma oscillators initiated by a single driven plasma oscillator under unidirectional interaction, we examine the effect of interaction strength on the response amplitude of each plasma oscillator. 
\begin{figure}[h]
\centering
\begin{minipage}{0.24\textwidth}
\centering 
\includegraphics[height=\textwidth,width=\textwidth]{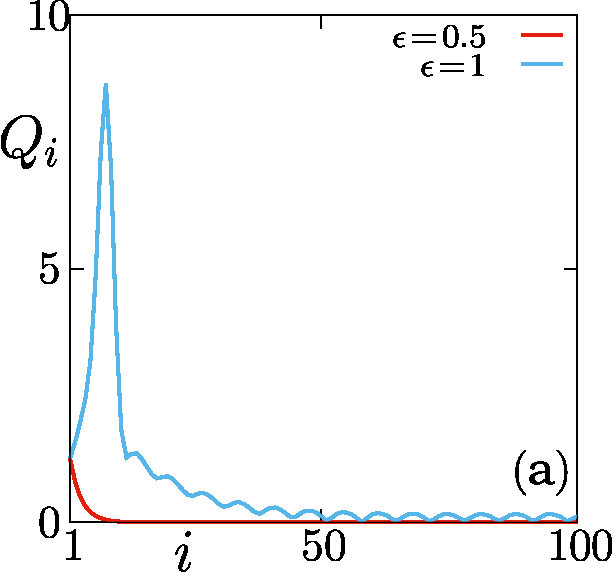}
\end{minipage}
\centering
\begin{minipage}{0.24\textwidth}
\centering 
\includegraphics[height=\textwidth,width=\textwidth]{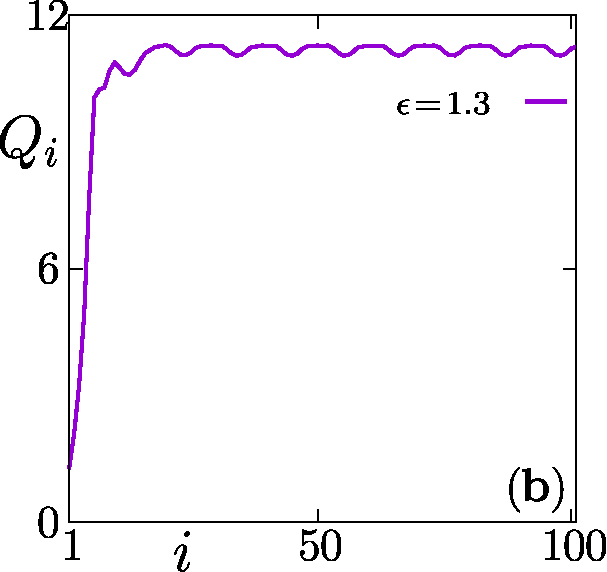}
\end{minipage}
\centering
\begin{minipage}{0.24\textwidth}
\centering 
\includegraphics[height=\textwidth,width=\textwidth]{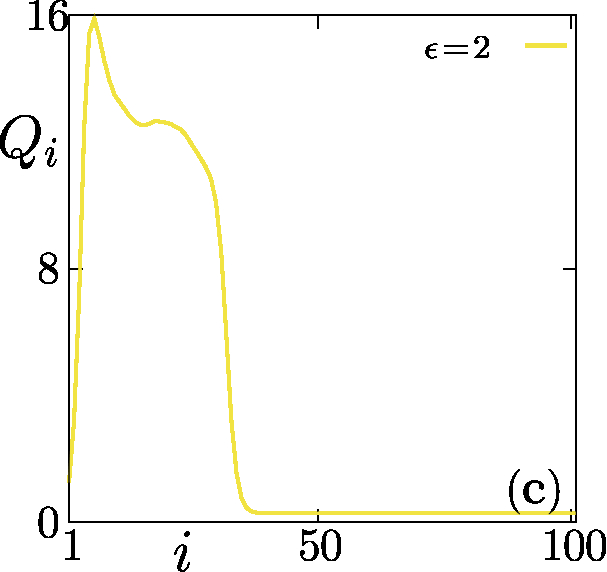}
\end{minipage}
\centering
\begin{minipage}{0.24\textwidth}
\centering 
\includegraphics[height=\textwidth,width=\textwidth]{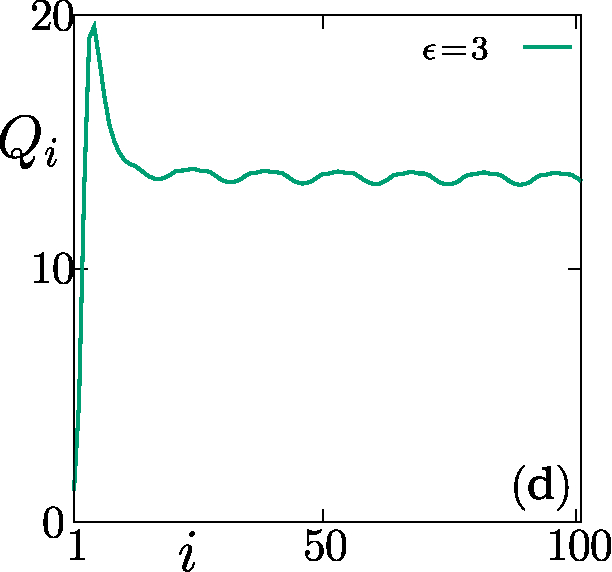}
\end{minipage}
\centering
\begin{minipage}{0.24\textwidth}
\centering 
\includegraphics[height=\textwidth,width=\textwidth]{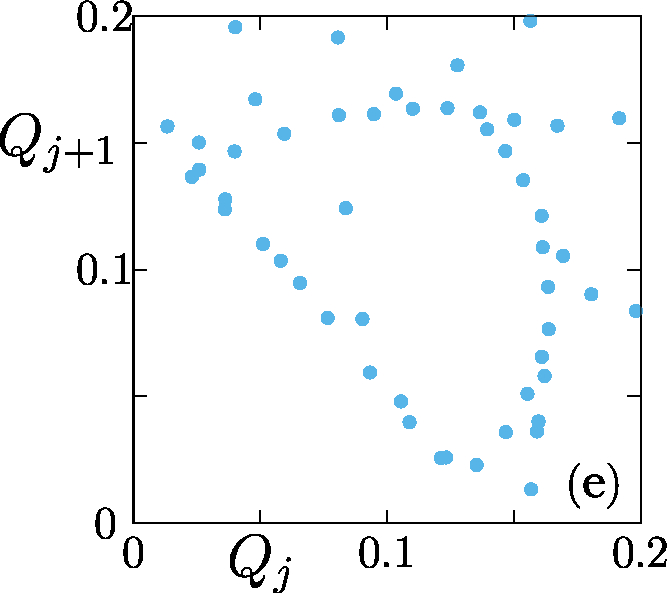}
\end{minipage}
\centering
\begin{minipage}{0.24\textwidth}
\centering 
\includegraphics[height=\textwidth,width=\textwidth]{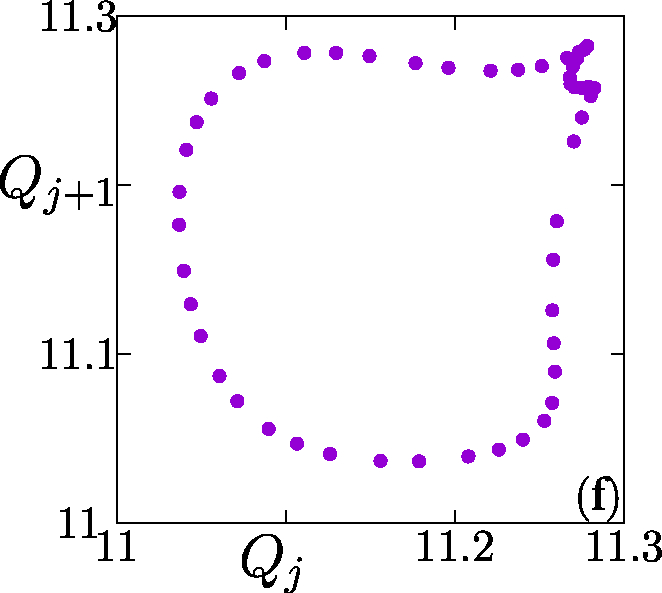}
\end{minipage}
\centering
\begin{minipage}{0.24\textwidth}
\centering 
\includegraphics[height=\textwidth,width=\textwidth]{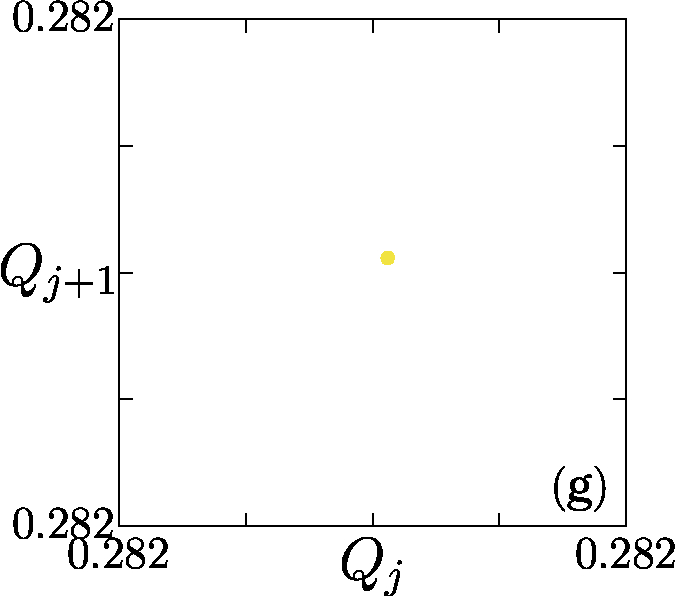}
\end{minipage}
\centering
\begin{minipage}{0.24\textwidth}
\centering 
\includegraphics[height=\textwidth,width=\textwidth]{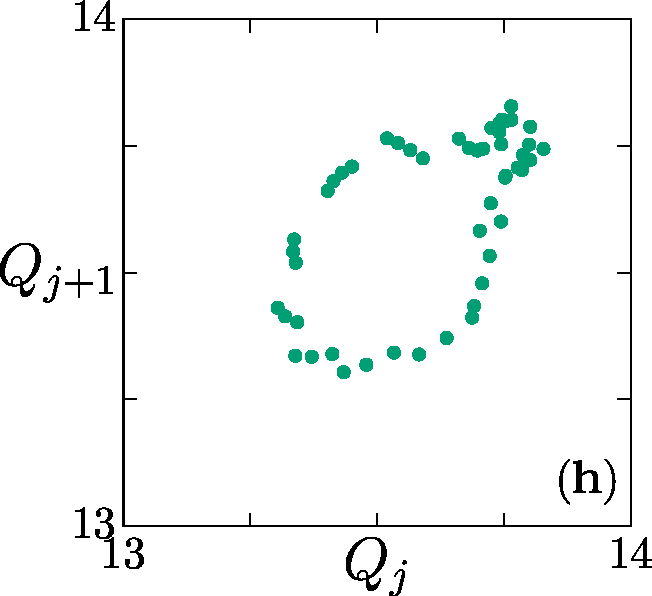}
\end{minipage}
\caption{(a-d) Signal transmission patterns and (e-h) amplitude return maps for different coupling strengths $\epsilon$.}
\label{figure6}
\end{figure}

For weak coupling strength $(\epsilon=0.5)$, the energized plasma oscillator effectively transmits signal only to a limited number of subsequent oscillators. Due to signal loss at each stage of transmission, the response amplitude diminishes rapidly beyond the $13$-th plasma oscillator. This indicates a strong attenuation of the signal as it propagates through the network of $n$-plasma oscillators. The significant signal loss is highlighted by the red curve in Fig.~\ref{figure6}(a). When the interaction strength increases to $\epsilon=1.0$, the energized plasma oscillator transmits signal more effectively, and each interacting plasma oscillator gains more energy than its immediate predecessor in the sequence. The blue curve in Fig.~\ref{figure6}(a) illustrates the signal transmission pattern corresponding to an interaction strength of $\epsilon=1.0$. In particular, the $7$-th plasma oscillator attains the highest amplification in the network, with a peak response amplitude of $Q_7=8.64$. Beyond this plasma oscillator, the signal is transferred with gradual losses at each subsequent interaction. This cumulative signal loss results in a progressive attenuation of response amplitudes. By the time the signal reaches the $30$-th plasma oscillator, the amplitude has dropped significantly to $Q_{30}=0.3$. After this point, the system no longer exhibits a monotonic amplitude decay; instead, the response amplitudes become irregular and non-monotonic, indicating complex response dynamics in the tail end of the network. This leads to a spatially heterogeneous response pattern, as the fluctuations between $0.02$ and $0.2$ observed from the $30$-th plasma oscillator onward. This behavior is shown in the amplitude return map shown in Fig.~\ref{figure6}(e), where scattered points are observed. The map is plotted using the response amplitudes of the plasma oscillators after the $49$-th in the network.

For $\epsilon=1.3$, the energized plasma oscillator transmits signal with successive gains across the network, as illustrated by the violet curve in Fig.~\ref{figure6}(b). Starting from the $19$-th plasma oscillator, the energy is sustained throughout the remainder of the network. The response amplitudes beyond this point fluctuate within a narrow range of $[11.02:11.3]$, forming a quasi-periodic structure. This behavior is further illustrated in the amplitude return map shown in Fig.~\ref{figure6}(f).

At $\epsilon=2.0$, signal is efficiently amplified at the beginning of the network, reaching a peak at the $5$-th plasma oscillator with a response amplitude of $Q_5=15.8$. Beyond this peak, the amplitude gradually attenuates, decreasing to a minimum of $Q=0.28$ near the $40$-th plasma oscillator. The remaining oscillators in the network transmit a weak, nearly constant amplitude, as indicated by the yellow curve in Fig.~\ref{figure6}(c). The amplitude return map is shown in Fig.~\ref{figure6}(g), reveals an isolated point, confirming a weak and nearly uniform strength signal transfer between the interacting plasma oscillators.

For strong interaction $(\epsilon=3.0)$, a sharp increase in response amplitude is observed within the first four plasma oscillators, with maximum amplification occurring at the $4$-th plasma oscillator, where $Q_4=19.54$ the highest recorded in the entire network. Beyond the $19$-th plasma element, the response amplitudes exhibit periodic modulations approximately every $14$ oscillator, forming a distinct wave-like spatial pattern. In this regime, the amplitudes fluctuate within a narrow range of $[13.3:13.8]$. This behavior is illustrated by the green curve in Fig.~\ref{figure6}(d). The corresponding return map is shown in Fig.~\ref{figure6}(h). The map exhibits a closed-loop pattern, characteristic of quasi-periodic signal transfer among the plasma oscillators. 

Figure~\ref{figure6} shows five different types of signal transmission patterns in the network, each occurring within a specific range of interaction strength $ \epsilon $, and is indicated by different color lines. For the red line signal transfer pattern ($ \epsilon \in [0,~0.75] $), the signal weakens quickly as it travels through the network, indicating the damped signal transmission. For the blue line signal transfer pattern,  region ($ \epsilon \in [0.755,~1.27] $), the amplitude is increased only at a few specific plasma oscillators, while the rest remain weak, and this is known as localized amplification at the specific plasma oscillator. For the violet line signal transfer pattern, region ($ \epsilon \in [1.275,~1.435] $), the signal is transmitted evenly across the entire network without loss, showing undamped transmission. For the yellow line signal transfer pattern, region ($ \epsilon \in [1.44,~2.93] $), the signal becomes stronger in the middle part of the network but then fades as it continues to propagate and we referred this region as intermediate signal or energy transmission. Finally, for the green line signal transfer region ($ \epsilon \in [2.935,~4.0] $), the signal reaches a peak at a certain oscillator and then continues to spread through the rest of the network with almost no loss, indicating a strong and undamped transmission. These transmission patterns highlight the complex nonlinear behavior of the network of plasma oscillators and show how the interaction strength $ \epsilon $ can be used to control how energy moves through the network (\ref{network_eqn}). 

An important feature in these transmission regimes is the emergence of interference-like patterns in the amplitude profiles, seen as uneven distributions centered around specific amplitude bands. This indicates complex dynamics in which each plasma oscillator is affected not only by the main driving frequency $\omega$ but also by additional frequencies generated through nonlinear interactions. These patterns result from nonlinear mixing between the external forcing and the plasma oscillator intrinsic response, causing spectral broadening and harmonic generation. Consequently, constructive and destructive interference shapes the spatial amplitude variations. To fully understand signal transmission, spectral analysis over a wide frequency range is necessary. This reveals dominant frequencies, harmonic interactions, and modulation effects, providing insight into energy transport and localization mechanisms in nonlinear oscillator networks. 
\section{Higher harmonic generation in $n$-plasma oscillators}
In this section, we investigate the generation of higher harmonics in a network of $n$ unidirectionally interacting plasma oscillators. To explore this behavior, we analyze the amplitude spectra for four representative interaction strengths: $\epsilon$ equals $0.5, 1, 1.3, 2, 3$ and $4$ as illustrated in Fig.~\ref{figure6}. 
\begin{figure}[h]
\centering
\begin{minipage}{0.4\textwidth}
\centering   
\includegraphics[width=\textwidth]{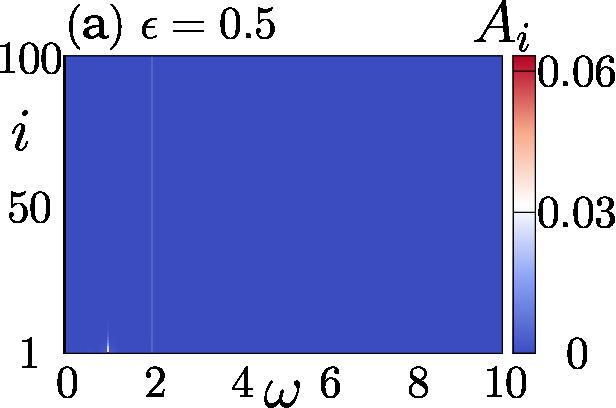}
\end{minipage}
\centering
\begin{minipage}{0.4\textwidth}
\centering  
    \includegraphics[width=\textwidth]{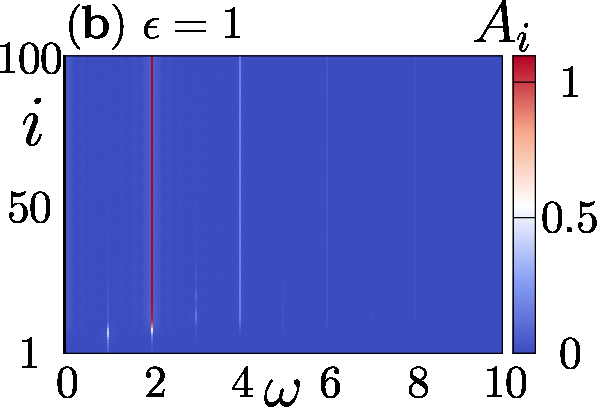}
\end{minipage}
\centering
\begin{minipage}{0.4\textwidth}
\centering  
    \includegraphics[width=\textwidth]{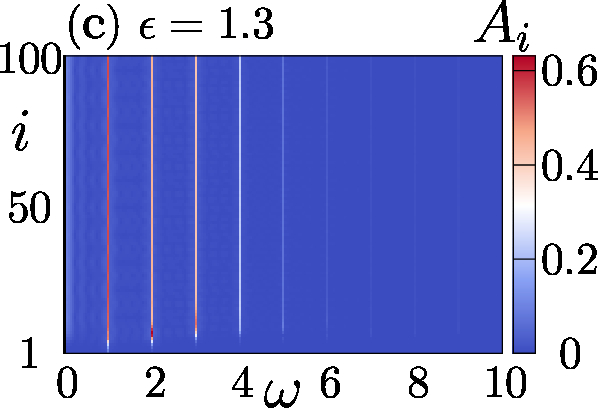}
\end{minipage}
\centering
\begin{minipage}{0.4\textwidth}
\centering 
    \includegraphics[width=\textwidth]{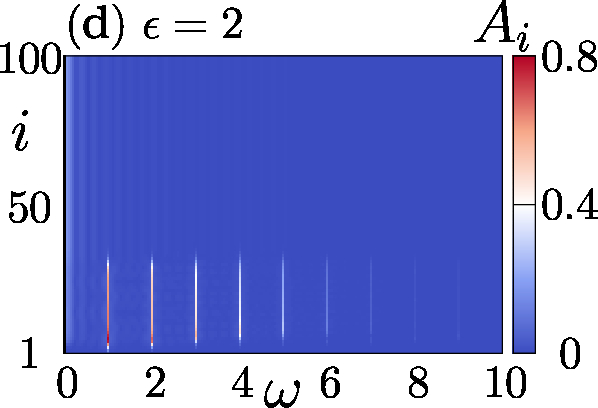}
\end{minipage}
\centering
\begin{minipage}{0.4\textwidth}
\centering 
    \includegraphics[width=\textwidth]{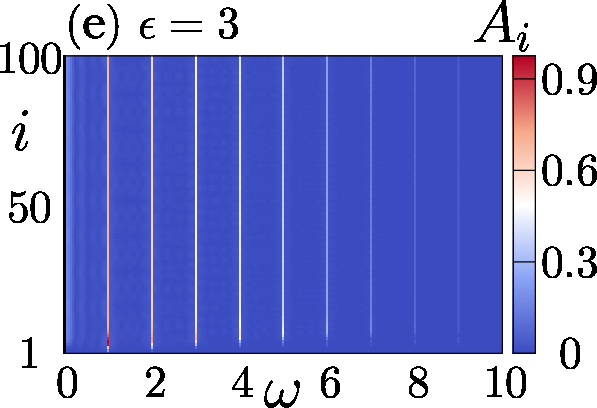}
\end{minipage}
\centering
\begin{minipage}{0.4\textwidth}
\centering 
    \includegraphics[width=\textwidth]{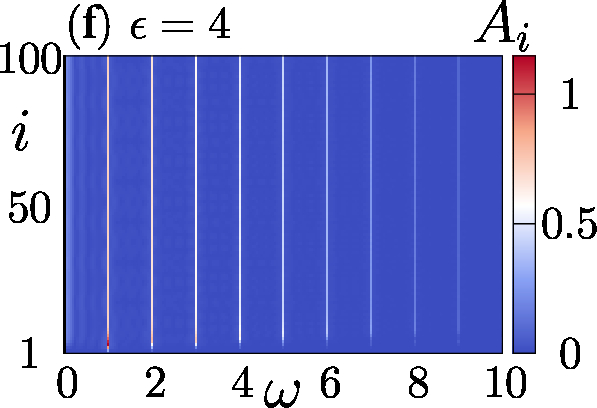}
\end{minipage}
\caption{Amplitude spectra for $100$ plasma oscillators for different coupling strengths $\epsilon$.}
\label{figure7}
\end{figure}

We begin our analysis with an interaction strength of $\epsilon=0.5$. Figure~\ref{figure7}(a) presents the amplitude spectra corresponding to this coupling value, which lies in the regime of signal (amplitude) attenuation $\epsilon\in[0:0.75]$. As seen in the figure, the spectral component at the fundamental frequency decreases progressively after the first few oscillators, indicating gradual suppression of the primary signal along the network. In contrast, the second-harmonic component persists and propagates throughout the network. This behavior reflects the emergence of a ghost resonance effect, wherein the second harmonic governs the transmission dynamics despite the attenuation of the fundamental mode. Notably, the isolated driven oscillator generates only a very weak second harmonic; however, once generated, this component is sustained and transmitted along the network. At $\epsilon=0.5$, the transmission is therefore dominated solely by the second harmonic.  

Figure~\ref{figure7}(b) shows the amplitude spectra for a coupling strength of $\epsilon=1.0$, which falls within the localized signal amplification regime 
$(\epsilon\in[0.755:1.27])$. In this regime, the driven signal undergoes initial amplification before gradually attenuating as it propagates along the network. A weak second harmonic component is generated at the first oscillator and subsequently enhanced and efficiently transmitted across the network. However, unlike the previous case, the present spectrum also exhibits weak odd harmonics that are strongly attenuated, while the even harmonics particularly the second harmonic remain prominent and propagate effectively. This behavior indicates that, at this coupling strength, the unidirectional interaction preferentially supports even-harmonic generation, leading to a dominant second harmonic response and selective suppression of odd order modes. 

The amplitude spectra for a coupling strength of $\epsilon=1.3$ is presented in Fig.~\ref{figure7}(c). From the spectra, it is evident that the fundamental frequency component of the input signal is strongly amplified and successfully transmitted across the entire oscillator network. This behavior consistently occurs within the interaction strength range $ \epsilon \in [1.275,~1.435] $, indicating that this regime is well-suited for efficient signal transmission. In this regime, all higher-order harmonics are also present but with progressively smaller amplitudes compared to the immediately preceding lower harmonics within each oscillator. This hierarchy of harmonic amplitudes indicates efficient fundamental frequency transmission accompanied by diminishing nonlinear contributions as the signal propagates through the coupled network.

In the regime $\epsilon\in[1.44:2.98]$, the signal undergoes strong amplification in the intermediate region of the network, followed by a gradual weakening toward the end of the network. To examine the evolution of the signal’s frequency content in this regime, Fig.~\ref{figure7}(d) displays the amplitude spectra for $\epsilon=2$. Similar to the previous case, multiple harmonic components are present; however, their amplitudes become progressively attenuated after the first few oscillators in the network.

For a coupling strength of $\epsilon=3.0$, the system enters a distinct signal transmission regime. In this case, the input signal introduced at the first plasma oscillator propagates through the entire network without attenuation. As discussed for Fig.~\ref{figure7}(c), this coupling strength leads to the generation of multiple harmonics in the subsequent oscillators along the network. After an initial amplification stage, the signal maintains an almost constant amplitude across the network for the higher harmonics. This behavior is clearly illustrated by the amplitude spectra shown in Fig.~\ref{figure7}(e). A similar transmission pattern persists for $\epsilon=4$, as shown in Fig.~\ref{figure7}(f).

Overall, increasing the coupling strength in the amplification regime enhances harmonic generation in successive oscillators. As the number of oscillators increases, the accumulation of higher harmonics leads to interference effects and increasingly complex signal waveforms in downstream elements. Since the harmonic balance method assumes a predominantly linear, single-frequency response, it cannot fully capture this complex multi-harmonic behavior, which explains the observed analytical–numerical mismatch in the strong amplification regime. 

\begin{figure}[h]
\centering
\begin{minipage}{0.3\textwidth}
\centering   
\includegraphics[width=\textwidth]{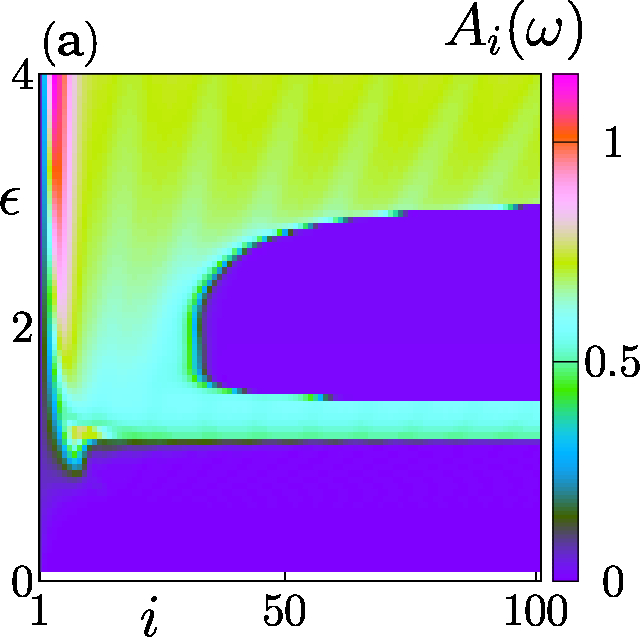}
\end{minipage}
\centering
\begin{minipage}{0.3\textwidth}
\centering   
\includegraphics[width=\textwidth]{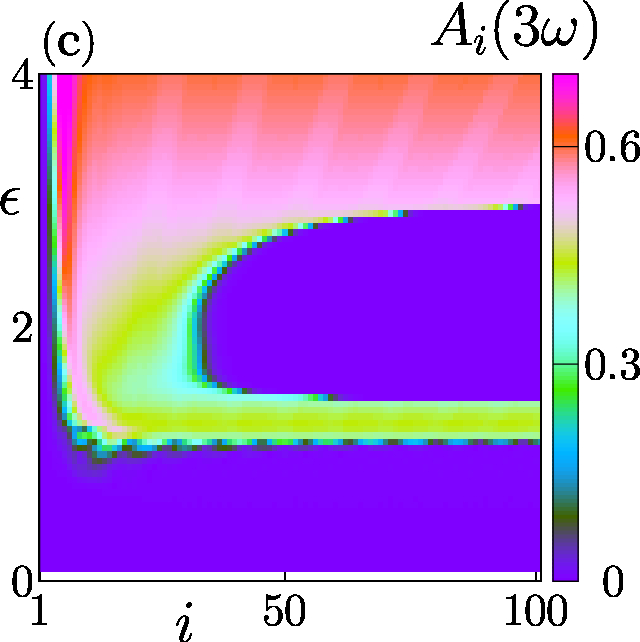}
\end{minipage}
\centering
\begin{minipage}{0.3\textwidth}
\centering   
\includegraphics[width=\textwidth]{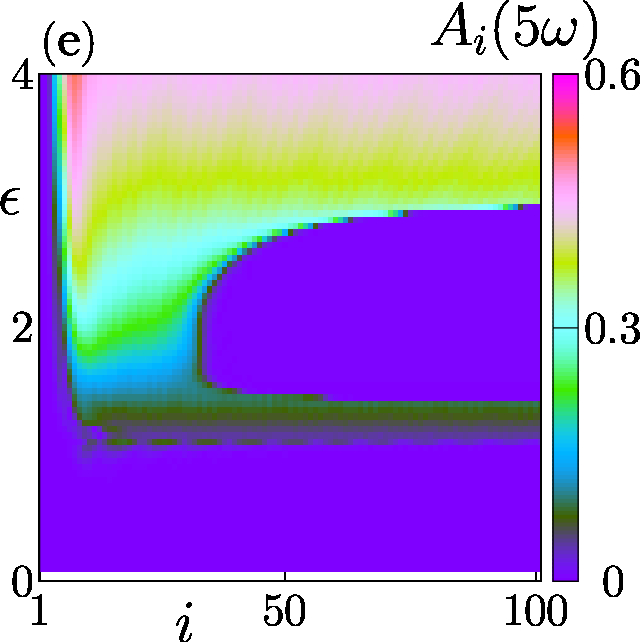}
\end{minipage}
\centering
\begin{minipage}{0.3\textwidth}
\centering   
\includegraphics[width=\textwidth]{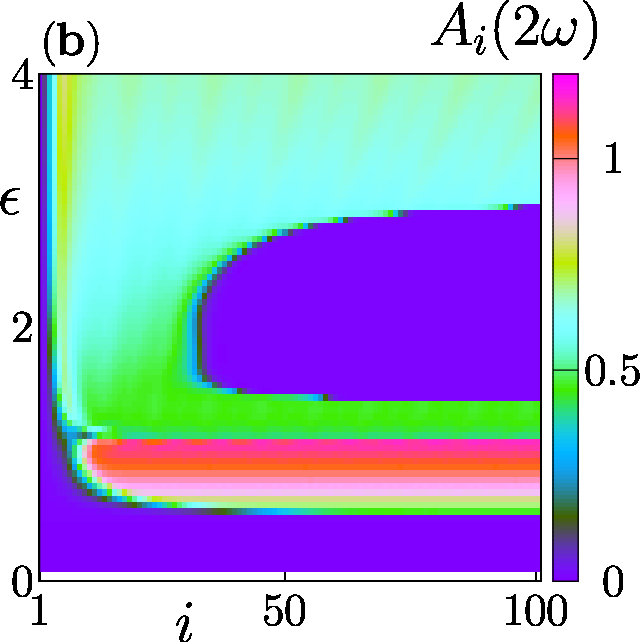}
\end{minipage}
\centering
\begin{minipage}{0.3\textwidth}
\centering   
\includegraphics[width=\textwidth]{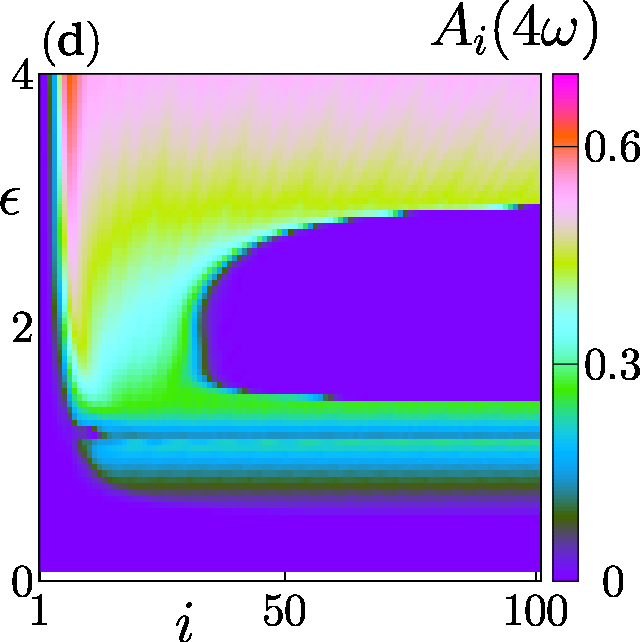}
\end{minipage}
\centering
\begin{minipage}{0.3\textwidth}
\centering   
\includegraphics[width=\textwidth]{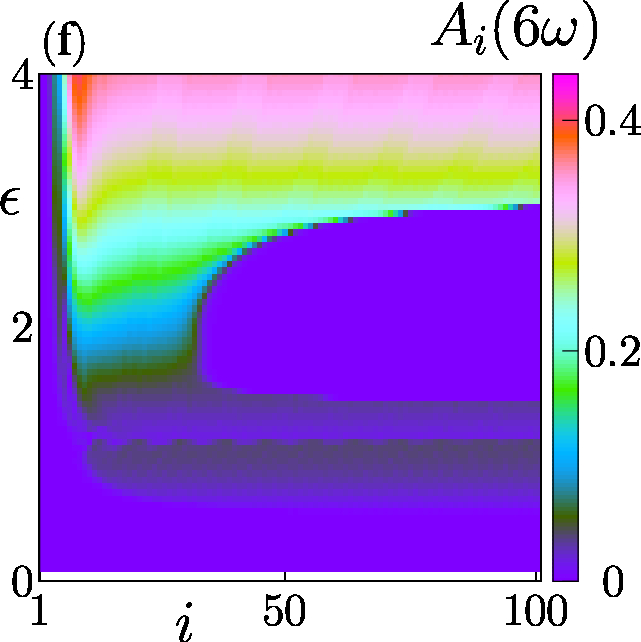}
\end{minipage}
\caption{Dependence of coupling strength $\epsilon$ on the fundamental and higher harmonic components in a unidirectionally coupled plasma oscillator network.}
\label{figure8}
\end{figure}
Figure~\ref{figure8} illustrates the dependence of the coupling strength $\epsilon$ on the fundamental and higher harmonic components of the transmitted signal in a unidirectionally coupled plasma oscillator network. Figure \ref{figure8}(a-f) shows the variation of the amplitude at the fundamental and higher harmonic frequencies $\omega,~2\omega,~3\omega,~4\omega,~5\omega$ and $6\omega$ respectively across $n$ plasma oscillator, Higher harmonics surface plots exhibit patterns analogous to those observed for the fundamental harmonic $\omega$, yet distinct variations in transmission behavior are evident for odd harmonics and even harmonics. 

Based on these observations, we find that the fundamental frequency $\omega$ and odd harmonics $3\omega$ and $5\omega$ exhibit five distinct transmission patterns, whereas even harmonics $2\omega$, $4\omega$ and $6\omega$ show four distinct transmission patterns. Interestingly, the ranges of the interaction strength $\epsilon$ corresponding to the different transmission regimes are nearly identical for even harmonics. In all cases, intermediate oscillators exhibit enhanced signal transmission, which is subsequently followed by damped propagation, localized amplification, and eventually undamped transmission as the coupling strength $\epsilon$ increases. These transmission behaviors consistently emerge in the unidirectionally coupled plasma oscillator network for higher values of $\epsilon$. Moreover, for coupling strengths $\epsilon>1.44$, the transmission regimes observed across different harmonic components remain approximately the same. 

\section{Role of System Parameters in Signal Propagation}

\begin{figure}[h!]
\centering
\begin{minipage}{0.6\textwidth}
\centering   
\includegraphics[width=\textwidth]{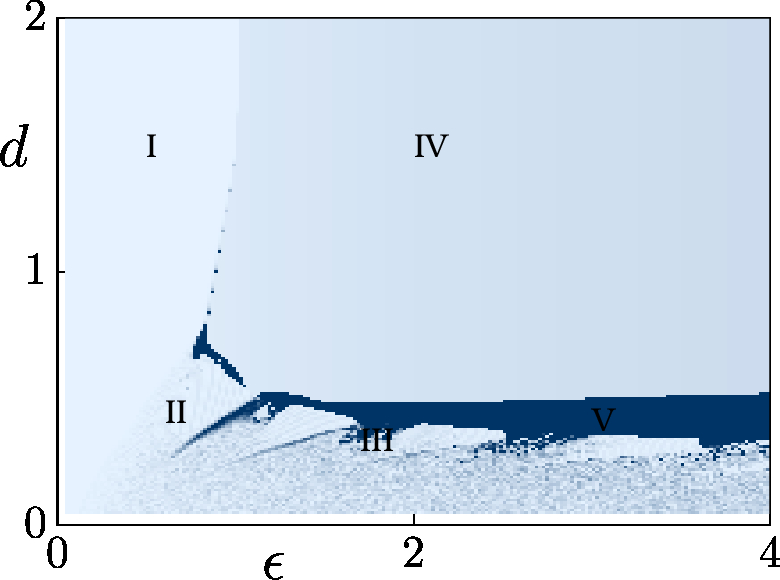}
\end{minipage}
\caption{Mapping of signal transmission dynamics with respect to the coupling parameters $d$ and $\epsilon$}
\label{figure9}
\end{figure}

The signal transmission behavior in a unidirectionally coupled system of plasma oscillators, modeled as \ref{network_eqn}), is strongly influenced by key system parameters. This section investigates the effects of the damping coefficient $ d $, the nonlinear asymmetry parameter $ \alpha $, and the cubic nonlinearity parameter $ \beta $ on the propagation dynamics. These parameters control the local response of individual oscillators and determine the overall amplitude, stability, and harmonic structure of the transmitted signal. By systematically varying these parameters under different coupling strengths $ \epsilon $, we aim to identify the conditions that promote or suppress effective signal transmission along the network.

Figure~\ref{figure9} presents the map of signal transmission corresponding to system~(\ref{network_eqn}). The different regions in the map represent distinct transmission behaviors:
Region~I corresponds to the damped signal transmission regime (red line in Fig.~\ref{figure6}(a));
Region~II represents localized amplification with damped transmission (blue line in Fig.~\ref{figure6}(a));
Region~III denotes undamped signal transmission (violet line in Fig.~\ref{figure6}(b));
Region~IV represents intermediate amplification followed by weakened transmission (yellow line in Fig.~\ref{figure6}(c));
and Region~V corresponds to localized amplification with sustained, undamped transmission (green line in Fig.~\ref{figure6}(d)). When $d>0.8$, variations in $\epsilon$ do not lead to undamped signal transmission; for all considered values of $d$ and $\epsilon$, the signal undergoes attenuation along the network. In contrast, for $d\in[0.28:0.72]$, varying $\epsilon$ gives rise to five distinct signal transmission regimes at the driving frequency along the network. As the results when nonlinear damping is very high or very less, however unidirectional coupling strength is the amplification of output signal of oscillator, no signal transmission.   
\begin{figure}[h!]
\centering
\begin{minipage}{0.4\textwidth}
\centering 
\includegraphics[height=\textwidth,width=\textwidth]{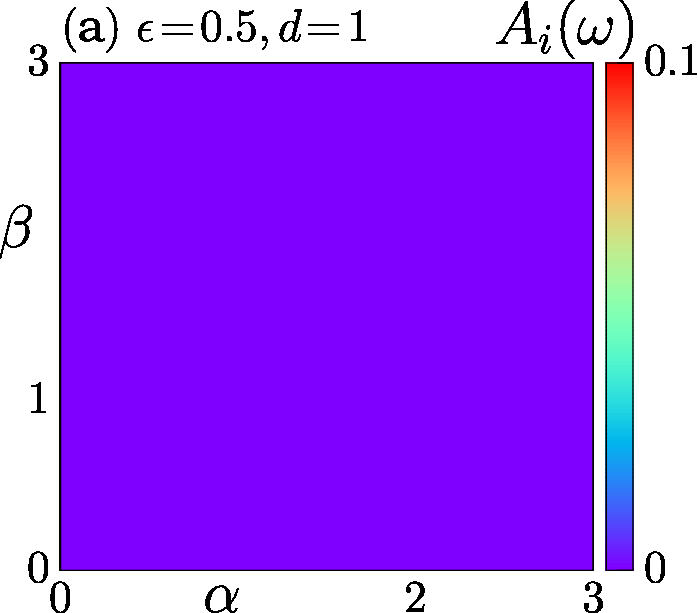}
\end{minipage}
\centering
\begin{minipage}{0.4\textwidth}
\centering 
\includegraphics[height=\textwidth,width=\textwidth]{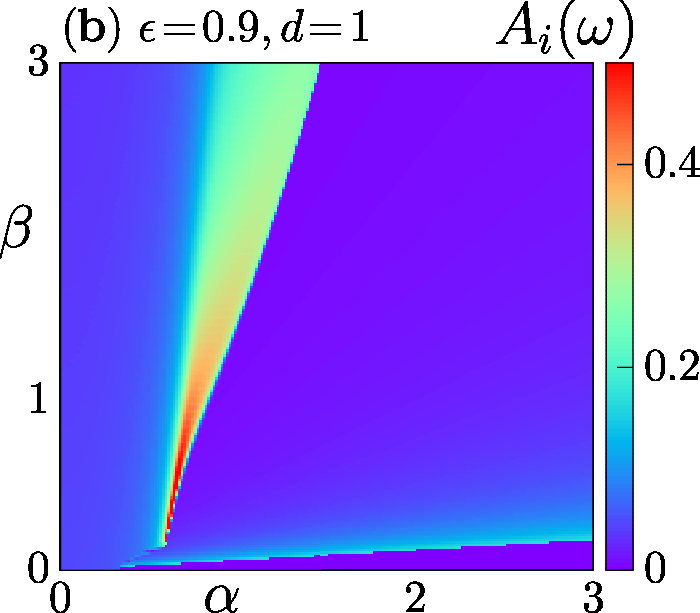}
\end{minipage}
\centering
\begin{minipage}{0.4\textwidth}
\centering 
\includegraphics[height=\textwidth,width=0.9\textwidth]{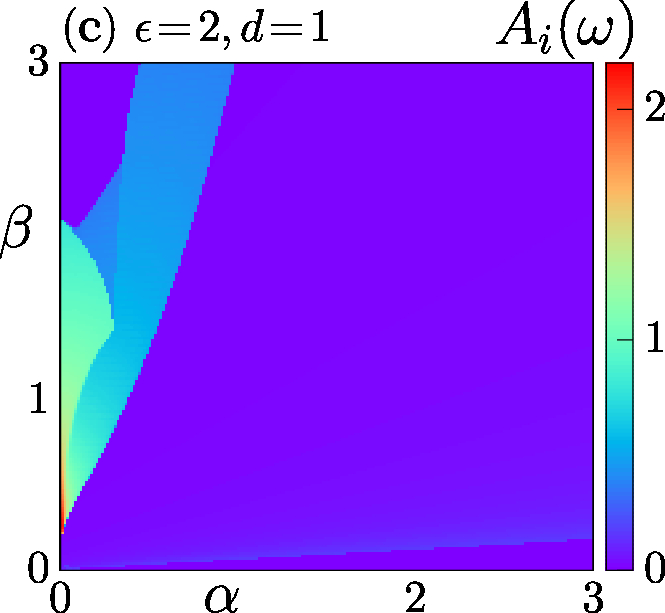}
\end{minipage}
\centering
\begin{minipage}{0.4\textwidth}
\centering 
\includegraphics[height=\textwidth,width=0.9\textwidth]{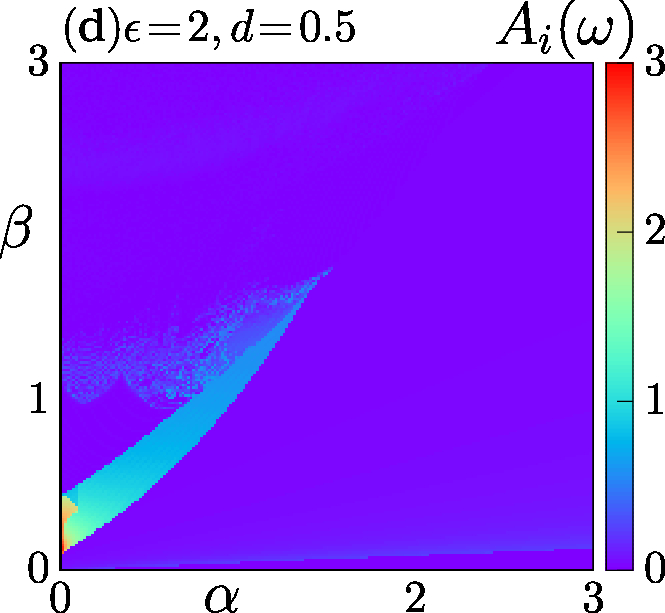}
\end{minipage}   
\caption{Amplitude profiles showing the influence of assymmetry and nonlinear parameters $\alpha$ and $\beta$ in unidirectionally coupled plasma oscillators.
}
\label{figure10}
\end{figure}

From the transmission map discussed above, we selected four representative combinations of the parameters $d$ and $\epsilon$ corresponding to Regions~I, II, III, and the boundary between Regions~I and II. For each selected parameter set, we constructed the amplitude surface plots in the $(\alpha,\beta)$ plane to analyze how these parameters influence the transmission dynamics.

In Fig.~\ref{figure10}(a), the chosen values of $\epsilon$ and $d$ correspond to the damped signal transmission region. Here, no significant variation is observed in the surface distribution of $\alpha$ and $\beta$, indicating that the system remains confined within the damped transmission regime. At the boundary between Regions~I and II, as shown in Fig.~\ref{figure10}(b), an optimal range of $\alpha$ and $\beta$ emerges, suggesting the onset of enhanced transmission. In contrast, for the parameter sets corresponding to Regions~II and III [Figs.~\ref{figure10}(c) and \ref{figure10}(d)], the transmission dynamics exhibit strong dependence on both $\alpha$ and $\beta$, implying that appropriate tuning of these parameters can effectively control and improve the signal propagation across the network. 

\section{Conclusion}
In this study, we investigate signal transmission in a unidirectionally coupled plasma oscillator network, with particular emphasis on harmonic generation, propagation characteristics, and the influence of system parameters. Our results demonstrate that the coupling strength $\epsilon$ plays a crucial role in controlling the evolution of the signal as it propagates through the network. For the fundamental frequency $\omega$ and odd harmonics, five distinct transmission profiles are identified, whereas four different patterns emerge for even harmonics. These transmission regimes include damped transmission, undamped propagation, intermediate amplification, and localized amplification, depending on the coupling strength.

As the externally driven signal propagates from one plasma oscillator to the next, second harmonics are generated at the driven oscillator, while higher-order harmonics emerge progressively due to the combined effects of coupling strength and the oscillator position within the network, for the chosen parameter set. Notably, higher-harmonic generation and strong amplification are found to be intrinsically linked: strong amplification occurs predominantly in regimes where higher harmonics are generated, and conversely, the emergence of higher harmonics is facilitated by strong amplification.

A detailed parametric analysis reveals that the damping parameter $d$ can switch the system between damped and undamped transmission regimes and significantly restrict the number of observable transmission patterns. The asymmetry parameter $\alpha$ modulates the overall spatial distribution of the signal, while the cubic nonlinearity parameter $\beta$ exhibits only a minimal influence within the investigated range.

These findings highlight the rich and tunable nature of signal transport in plasma oscillator networks and provide important insights into the control of energy and information flow in nonlinear media. The present results are obtained for a deterministic feedforward plasma oscillator network; investigating the influence of stochastic perturbations on harmonic propagation and transmission regimes remains an interesting direction for future study.

\section*{Acknowledgments}
This work has received financial support by the Spanish State Research Agency (AEI) and the European Regional Development Fund (ERDF, EU) under Project No. PID2023-148160NB-I00 (MCIN/AEI/10.13039/ 501100011033).

\bibliography{zbibliography} 

\end{document}